\documentclass[prl,aps,twocolumn,superscriptaddress,showpacs, amsmath,
amssymb]{revtex4} 
\usepackage{graphicx}
\usepackage{psfrag}
\usepackage{color}
\usepackage{subfigure}
\definecolor{dred}{rgb}{0.7,0.0,0.0}

\newcommand{\e}{\textrm{e}}

\begin{document}

\title{Intrinsic coupling of orbital excitations to spin fluctuations in Mott
   insulators}

\author { Krzysztof Wohlfeld }
\affiliation{ IFW Dresden, P. O. Box 27 01 16, D-01171 Dresden, Germany}

\author { Maria Daghofer }
\affiliation{ IFW Dresden, P. O. Box 27 01 16, D-01171 Dresden, Germany}

\author { Satoshi Nishimoto }
\affiliation{ IFW Dresden, P. O. Box 27 01 16, D-01171 Dresden, Germany}

\author { Giniyat Khaliullin }
\affiliation{ Max-Planck-Institut f\"ur Festk\"orperforschung,
              Heisenbergstrasse 1, D-70569 Stuttgart, Germany }

\author { Jeroen van den Brink }
\affiliation{ IFW Dresden, P. O. Box 27 01 16, D-01171 Dresden, Germany}

\date{\today}

\begin{abstract}
We show how the general and basic asymmetry between two fundamental degrees
of freedom present in strongly correlated oxides, spin and orbital, has very
profound repercussions on the elementary spin and orbital excitations. Whereas 
the magnons remain largely unaffected, orbitons become inherently coupled 
with spin fluctuations in spin-orbital models with antiferromagnetic and 
ferroorbital ordered ground states. The composite orbiton-magnon modes that 
emerge fractionalize again in one dimension, giving rise to spin-orbital 
separation in the peculiar regime where spinons are faster than orbitons. 
\end{abstract}


\maketitle

In transition metal oxides, different $3d$ orbitals near the Fermi level can
have similar energy and thereby contribute to the low-energy physics. In
presence of strong correlations charge fluctuations become
suppressed and a Mott insulator is realized when there is a commensurate
number of electrons per unit cell. The effective Hamiltonian that emerges, 
often referred to as Kugel-Khomskii (KK) Hamiltonian, can be expressed in 
terms of spin and orbital operators~\cite{KK1982}. From a formal viewpoint 
the spin and orbital operators are very similar because they form identical 
algebras, but the way in which these operators enter into realistic KK 
Hamiltonians is very different. 

While the spin wavefunction is in essence rotationally invariant and the
coupling between spin operators therefore $SU(2)$ symmetric, the symmetry in
orbital space is much lower due to 
the ubiquitous crystal field acting on the orbital wavefunctions in crystals.
This reduced
symmetry causes the orbital-orbital interaction to be very anisotropic in
space and often inherently frustrated, which causes exotic effects such as
macroscopic degeneracy of the groundstate or the emergence of non-Abelian
topological excitations in the case of the compass or Kitaev models,
respectively~\cite{Jackeli:2009p2016}, which may be relevant for quantum
computation~\cite{Kitaev20032,Doucot:2005p2466}. We will show that this
asymmetry between spin and orbital degrees of freedom has fundamental
repercussions on the coupling of the elementary spin and orbital excitations
-- the magnons and orbitons, respectively: whereas the magnons remain largely 
unaffected, orbitons become inherently coupled with spin fluctuations in 
spin-orbital models with antiferromagnetic (AF) and ferroorbital (FO) ordered 
ground states. This is relevant in the experimental context as substantial 
progress is being made in measuring orbitons~\cite{Ulr06, Ulr09} and their 
dispersion, in particular in resonant inelastic x-ray scattering 
(RIXS)~\cite{Kot01, Ame11}, where we predict the coupling of the orbiton to 
magnetic fluctuations to be clearly discernible. 

In the \emph{standard} approach the complex problem of intertwined
spin-orbital excitations is solved using a mean-field decoupling, i.e. 
considering magnons in a fixed orbital background or orbitons in a fixed spin 
background. While such an approach can work well to obtain the correct spin 
and orbital orderings consistent with the Goodenough-Kanamori 
rules~\cite{KK1982}
%
and in some cases with ferromagnetic (FM) order 
\cite{vdB98},
we show here that it fails to describe orbital 
excitations even qualitatively correctly for a number of spin-orbital 
models. We avoid this decoupling by mapping the coupled orbiton-magnon 
dynamics onto the well-controlled problem of a hole propagating in a magnetic 
background: 
the extensively studied single hole $t$--$J$ model. 
In particular, we find that in one dimension (1D) the orbital excitation 
fractionalizes into freely propagating spinon and orbiton, giving rise to 
spin-orbital separation in the peculiar regime where spinons are faster 
than orbitons.

{\it Model and problem statement.---} 
The generic form of the KK Hamiltonian~\cite{KK1982} in the Mott-insulating 
limit is
\begin{align}\label{eq:h}
{\mathcal H} = 
2J \Bigl( \sum_{\langle i,j\rangle} H^S_{ i,j} H^T_{ij}  + \sum_{i} H^T_{i} \Bigr),
\end{align}
where $i$ and $j$ are lattice sites and on each bond 
$\langle i,\!j \rangle $
the spin-spin interaction is $H^S_{ij}$ whereas the orbital-orbital
one is $H^T_{ij}$. 
To capture the {\it generic} differences between orbitons and magnons, 
it is enough to break the SU(2) rotation symmetry in orbital space,
which is here achieved by the large local crystal field breaking the degeneracy between the
orbitals, expressed as $H^T_{i} = \frac{E_z}{2J} T^z_i$. We
keep for simplicity the rotational symmetry in the interactions so that for the spins 
$H^S_{ij}={\bf S}_{i}\cdot{\bf S}_{j}+\tfrac{1}{4}$ and for the orbitals
$H^T_{ij}={\bf T}_{i}\cdot{\bf T}_{j}+\tfrac{1}{4}$, where 
${\bf S}$ (${\bf T}$) 
are the spin (orbital) operators that fulfill the $SU(2)$ algebra for 
$S=1/2$ ($T=1/2$) spins (pseudospins). The constant $J>0$ 
gives 
the energy 
scale of the spin-orbital superexchange and the symmetry breaking field for 
the orbitals is $E_z$. 
Note that for the 
case of $E_z=0$ 
(not considered here), this model has an $SU(4)$ symmetry even higher than the
combined $SU(2)\times SU(2)$ symmetries, which results in the ground state
given by the Bethe Ansatz and composite spin-orbital gapless excitations 
in addition to the separate spin and orbital ones (see, e.g., 
Ref.~\cite{Li99}). 
This model describes the low-energy physics, determined by
singly occupied sites, of a two-orbital Hubbard model in the limit of
a large onsite Coulomb repulsion $U$ and vanishing Hund's exchange
$J_H$, cf. Eqs. (1-5) and (29) in Ref. \cite{vdB98}.

Here we are interested in the orbital excitations of the model 
Eq.~(\ref{eq:h}) when $E_z \gg J$ i.e., the orbital splittings are larger
than magnetic coupling energy.
This is a realistic regime for many strongly correlated compounds such 
as 1D or two-dimensional (2D) cuprates \cite{note2}. 
As one degree of freedom is completely polarized, the ground state is
easily found and given by all
electrons occupying a single orbital in an AF state. 
But while decoupling spin and orbital degrees of freedom 
works here for the ground state, it is not at all appropriate 
for \emph{orbital} \emph{excitations} -- this is the
problem investigated below.

{\it Decoupling of spin and orbital sector.---}  The ground state $| \psi
\rangle = | \psi_S \rangle\otimes| \psi_O \rangle$ of Eq.~(\ref{eq:h}) is 
described by the ground state $| \psi_S \rangle =|\textrm{AF}\rangle$ 
of an AF Heisenberg system formed by spins in the lower-energy
orbital, i.e., an FO ordered state $| \psi_O \rangle =|\textrm{FO}\rangle$. 
The orbital excitations are reached by flipping an orbital, i.e., by promoting 
an electron at site $j$ from the occupied lower orbital to the empty 
higher band at the same site, expressed by the orbital raising operator 
$T^+_j$. The momentum-dependent orbital excitation is given by 
$T^+_k=\sum_j \e^{i  kj}T^+_j$ [$T^-_k=(T^+_k)^\dag$], and the
spectral function describing its dynamics is  
\begin{align} \label{eq:spectral}
O(k, \omega)=\frac{1}{\pi} \lim_{\eta \rightarrow 0} 
\Im \langle \psi | T^-_{k} 
\frac{1}{\omega + E_{\psi}  - \mathcal{{H}} -
i \eta } T^+_k | \psi \rangle.
\end{align}

\begin{figure}[t!]
\includegraphics[width=0.48\textwidth,trim = 30 25 50 40,
clip]{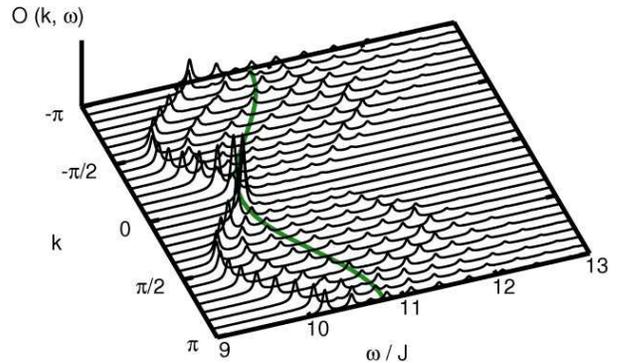}\label{fig:Ak_t1} 
\caption{(color online)
Spectral function $O(k, \omega)$ of orbital excitation obtained via the 
mapping onto the $t$--$J$ model, Eq. (\ref{eq:spectraltJ}), evaluated 
using Lanczos exact diagonalization on a 28 site chain. A broadening 
$\eta = 0.03J$ and $E_z = 10J$. The thick
line shows orbital 
excitation in a mean-field (orbital-wave) approach, Eq.~(\ref{eq:spectralow}). 
}
\label{fig:Ak_map}
\end{figure}

First we discuss the orbiton spectral function 
by taking  the orbital excitation
to be independent of the magnetic excitation. One can then rewrite the
orbital operators by use of Holstein-Primakoff bosons 
(see, e.g., Ref.~\cite{Wohlfeld:2009p2187}), keep only quadratic terms in the 
expansion (orbital-wave theory) and, noticing that the ground state 
$|\psi_O \rangle$ does not contain bosons, obtain 
\begin{align}
\label{eq:spectralow}
O (k, \omega)= \delta [\omega - \omega_{OW}(k)], 
\end{align}
with a mean-field orbital-wave dispersion 
$\omega_{OW}(k)=E_z - \frac{1}{2}zJ_{\rm OW}(1-\gamma_k)$.
Here, $z$ is the coordination number, $\gamma_k$ is the lattice structure factor, and the 
effective orbital exchange constant $J_{\rm OW} = 2J \langle \psi_S | 
{\bf S}_{i } \cdot {\bf S}_{j} + \frac{1}{4} | \psi_S \rangle.$
The orbital excitation on the mean-field level is thus a quasiparticle with a
cosine-like dispersion with period $2 \pi$: for example in 1D  
we obtain an effective reduced $J_{\rm OW} \simeq -0.4 J\ll J$, cf. the
thick line in Fig.~\ref{fig:Ak_map}. 
%
\begin{figure}[t!]
\includegraphics[width=0.48\textwidth]{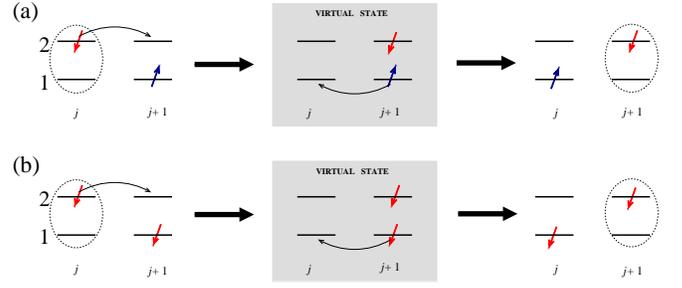}
\caption{(color online)
Two superexchange processes moving an electron in an excited orbital 
(indicated by oval) from site $j$ to its neighbor $j+1$:
(a) and (b) describe orbiton motion when spins along the bond are
antiparallel or parallel, respectively, see text.
The states in the grey middle panels are not
part of the low-energy Hilbert space corresponding to
(\ref{eq:h}). These virtual excitations within the full two-orbital
\emph{Hubbard} model illustrate the origin of 
those superexchange interactions
of Hamiltonian (\ref{eq:h}) that propagate the orbiton;
note that the spin of the excited electron is
conserved.
}
\label{fig:SE}
\end{figure}
An analogous procedure for {\it magnons} in FM planes with
alternating orbitals (AO) has been applied to LaMnO$_3$ or KCuF$_3$, and
similarly yields magnons with a reduced bandwidth, but without any
other trace of the AO order, in agreement with
experiment~\cite{Mou96}. We will see, however, that for orbitons this
framework of mean-field decoupling is greatly oversimplified. 

{\it Mapping onto an effective $t$--$J$ model.---}
The orbital-wave approximation puts all the impact of the AF order 
into $J_{\rm OW}$, which is a mean-field average of the sum of two distinct
superexchange processes in which the orbital excitation may propagate 
through the lattice. The first one, which corresponds to $(T^+_j T^-_{j+1}+h.c.)(S^+_j S^-_{j+1} + h.c.)$ 
processes in Hamiltonian (\ref{eq:h}), allows for orbiton propagation
when the spins on the bond are antiparallel, see Fig.~\ref{fig:SE}(a).
The second one, which corresponds to $(T^+_j T^-_{j+1}+h.c.)(S^z_j S^z_{j+1} + 1/4)$ 
processes in Hamiltonian (\ref{eq:h}), allows for orbiton propagation
when the spins on the bond are parallel, see Fig.~\ref{fig:SE}(b).
Crucially (see next paragraph), this figure illustrates that in both cases the spin of the electron in the upper
orbital 2 is conserved during the orbiton propagation. This is because spin-orbital Hamiltonian (\ref{eq:h}) is, 
as mentioned above, a low energy limit of the two-orbital Hubbard model with the Hund's exchange $J_H= 0$ 
and the spins of individual electrons in the superexchange process cannot be flipped
(see middle panels of Fig.~\ref{fig:SE}). 
In more realistic spin-orbital models the Hund's exchange is finite
\cite{KK1982}, but the processes which would not conserve the electron's
spin in the excited orbital are small ($\propto J_{H}/U$) and thus could
be neglected.

To be explicit, we now focus on the 1D case and employ a Jordan-Wigner 
transformation~\cite{Jordan:1928p2583}. 
2D and three-dimensional (3D) cases
are discussed
afterwards. We thus introduce 
$
S^+_j  = \beta_j  \e^{  i \pi Q }, \
S^-_j  =  \e^{ - i \pi Q } \beta^\dag_j, \
S^z_j = \frac12 - n_{j \beta}, 
$
where $Q= \sum_{l=1}^{j-1} n_{l \beta}$ and $\beta^\dag_j$ create spinons while 
$
T^+_j = \e^{ - i \pi \bar{Q} } \alpha^\dag_j, \
T^-_j  = \alpha_j \e^{  i \pi  \bar{Q}}, \
T^z_j  = n_{j \alpha} - \frac12,
$
where $\bar{Q} = \sum_{l=1}^{j-1} n_{l \alpha}$ and $\alpha^\dag_j$ creates 
a pseudospinon. Since the spin of the propagating electron in the upper 
orbital 2 is conserved (see above), one may calculate the spectral  
function $O(k, \omega)$ for, e.g., spin-up in the upper orbital. 
We are then allowed to replace 
$T^{\dag}_k \rightarrow \sum_{j} \e^{ikj} T^+_j (\frac12 + S^z_j) = 
\e^{ - i \pi \bar{Q} }\alpha^\dag_j (1- n_{j \beta})$ in Eq.~(\ref{eq:spectral})
and thus terms in the Hamiltonian that create or annihilate both a spinon and a
pseudospinon at the same site lead to a vanishing contribution to the
spectral function. 
Phase 
factors cancel in one dimension and we obtain
\begin{align} \label{eq:spectralpl}
O(k, \omega)=\frac{1}{\pi} \lim_{\eta \rightarrow 0} 
\Im \langle \bar{\psi} | \alpha_k \frac{1}{\omega \! +\! E_{\bar{\psi}}  \!-\!
\mathcal{\bar{H}}\! -\! E_z \! -\! i \eta } \alpha^+_k | \bar{\psi} \rangle, 
\end{align}
with the effective fermionic Hamiltonian 
\begin{align} \label{eq:eff}
\mathcal{{\bar{H}}}\! =&\! -\frac12  J\! \sum_{\langle i, j \rangle}
\big(\beta^\dag_i\alpha_i\alpha^\dag_{j} \beta_{j}\! +\! \alpha_i
\alpha^\dag_{j} \!+\! h.c.\big)\! 
\nonumber \\ 
&\!+\! J\! \sum_{\langle i, j \rangle} \Big[ \frac12 (\beta^\dag_i \beta_{j} \!+\! h.c.)  
\!-\! \frac12 (n_{i \beta}\! +\! n_{j \beta} )  \! +\! n_{i \beta} n_{j \beta} \Big],
\end{align}
and an implicit constraint $\forall_j \  \beta^\dag_j \beta_j + \alpha^\dag_j
\alpha_j \leq 1$. Here $|\bar{\psi}\rangle$ is a tensor product of the 
magnetic ground state $|\psi_S\rangle=|\textrm{AF}\rangle$ expressed in 
terms of spinons and a vacuum state for pseudospinons [recall that we 
consider here a single orbiton only; this also allowed us to skip 
quartic terms in pseudospinons in Eq. (\ref{eq:eff})]. 

At this point, we observe that the resulting effective Hamiltonian is in fact
a Hamiltonian for the $t$--$J$ model written in terms of the Jordan-Wigner
fermions with the above constraint \cite{Barnes:2002p2577}. By introducing
the electron operators 
${p}_{j \uparrow} = \alpha^\dag_j$, $\ {p}_{j \downarrow} = \alpha^\dag_j
\beta_j \e^ { i \pi Q }$ 
acting in the restricted Hilbert space without double occupancies we obtain 
\begin{align} \label{eq:spectraltJ}
O(k, \omega)\!=\!\frac{1}{\pi} \lim_{\eta \rightarrow 0} 
\Im \langle {\tilde{\psi}}  |{p}^{\dag}_{k \uparrow}   \frac{1}{\omega \!+\!
E_{{\tilde{\psi}} }  \!-\! \mathcal{{\tilde{H}}} \!-\! E_z \!-\! i \eta } {p}_{k
\uparrow}  |{\tilde{\psi}} \rangle, 
\end{align}
with the $t$--$J$ Hamiltonian 
\begin{align} \label{eq:tJ}
{\mathcal {\tilde{H}}}\! = &\! -\! t \!\sum_{\langle i, j \rangle, \sigma} ({p}^\dag_{i \sigma}
{p}_{j \sigma}\! +\! h.c.) 
\! +\! J \sum_{\langle i, j \rangle } ({\bf S}_{i } \cdot {\bf S}_{j} \! +\! \frac14 {n}_{i}
{n}_{j}), 
\end{align}
where ${n}_{j } =\sum_{\sigma} n_{jp\sigma}$ and the hopping parameter $t$ is 
defined as $t=J/2$~\cite{note}. The ground state $|\tilde{\psi}\rangle$ is now
the tensor product of a vacuum state for holes and the $|\psi_S\rangle$ state. 
We have thus mapped the {\it single orbiton} in the FO and AF chain, with
dynamics governed by Hamiltonian (\ref{eq:h}), onto a {\it single hole} doped 
into the undoped AF chain with its dynamics governed by Hamiltonian 
(\ref{eq:tJ}). 

{\it Numerical results for the $t$--$J$ model.---} 
To flesh out the resulting coupling between orbitons and spin fluctuations we
use Lanczos exact diagonalization to evaluate Eq. (\ref{eq:spectraltJ}) on a
finite chain (28 sites).  The spectral function is shown in
Fig.~\ref{fig:Ak_map}: the spectrum 
differs qualitatively from the
orbital-wave result shown as a thick 
line in Fig.~\ref{fig:Ak_map}. It now
consists of multiple peaks (expected to merge into incoherent spectrum in the
thermodynamic limit) instead of one single excitation. There is a dominant
feature at the lower edge of the spectrum, but its periodicity is $\pi$ 
reflecting the doubled unit cell of the AF order.  

\begin{figure}[t!]
\includegraphics[width=0.48\textwidth]{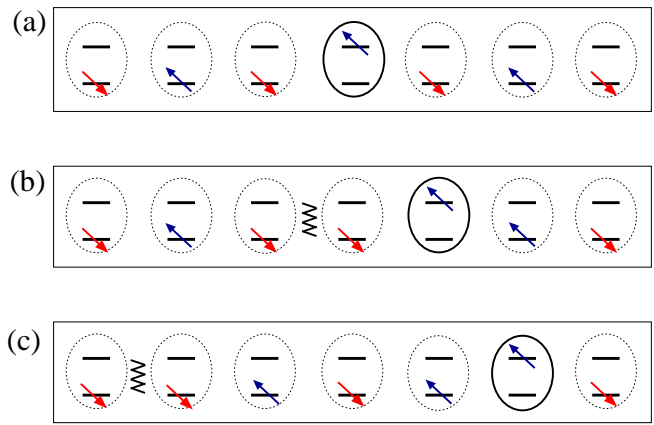}
\caption{(color online)
Schematic representation of the orbital motion and the induced spin 
fluctuations giving rise to spin-orbital separation in 1D. The first hop 
of the excited state (a $\to$ b) creates a spinon (wavy line) that moves via 
spin exchange $\propto J$. Next hop (b $\to$ c) does not produce any extra 
spinons: an `orbiton' freely propagating as a `holon' with an effective 
hopping $t\sim J/2$ is created.
}
\label{fig:4}
\end{figure}

The $t$--$J$ model with $J>t$~\cite{note} is not easily accessible in the Hubbard type
models, as it would formally correspond to small onsite interaction $U$, 
where the $t$--$J$ model is no longer valid. In this regime, the spinon moves 
faster than the holon, and the entire lower edge of the spectrum is thus given 
by `holon' states~\cite{Brunner:2000p2579}. If the orbiton takes the place 
of the holon as argued here, this exotic behavior should be observable in 
RIXS experiments and one would expect a dominant excitation 
with orbiton-character with dispersion $\omega\approx E_z-2t \sin |k|$ at the 
bottom of the spectrum. The latter would be expected to extend up to 
$\omega\approx E_z+ \sqrt{J^2+4t^2 - 4tJ \cos k } $ and to contain an 
intermediate feature still well-visible within the continuum with the 
dispersion of the purely orbiton-character scaling as 
$\omega\approx  E_z+2t \sin |k|$.

Figure~\ref{fig:4} illustrates how electron exchange processes can let an 
orbital excitation propagate through the system after creating a spinon in 
the first step. The spinon itself moves via spin flips $\propto J >t$, 
faster than the orbiton, and the two get well separated. The orbital-wave
picture, on the other hand, would require the orbital excitation to move 
without creating the spinon in the first step. As can be inferred 
from Fig.~\ref{fig:4}, this is only possible for
imperfect N\'eel AF spin order so that the \emph{averages} of processes shown in
Fig.~\ref{fig:SE}(a) and (b) are finite.

{\it 2D and 3D.---}
Remarkably, 
%
the standard OW picture becomes even worse in higher
dimensions: 
in 2D (3D) cubic lattices, the mean-field orbital coupling $J_{\rm 
OW}$ almost vanishes due to $\langle \psi_S|{\bf S}_{i }\cdot{\bf S}_{j} + \frac{1}{4} |\psi_S \rangle
\simeq -0.08\ (-0.05)$; the
orbital dynamics is then entirely governed by coupling to the spin
fluctuations~\cite{Kha00}. 
The $t$--$J$ model description of a \emph {single} orbiton Eqs. (\ref{eq:spectraltJ}-\ref{eq:tJ}) is 
general and valid for any dimension: the mapping rests entirely on the fact 
that spin on the excited orbital is conserved. In particular, the Jordan-Wigner 
fermionization applied to the 2D case gives the $t$--$J$ model expressed in terms
of Jordan-Wigner fermions~\cite{Barnes:2002p2577} 
and consequently the 2D version of Eqs. (\ref{eq:spectraltJ}-\ref{eq:tJ}),
see \cite{supp} for details.

A very good approximate solution for the higher-dimensional $t$--$J$
model with $J>t$ can be obtained by perturbation theory, because $J>t$
corresponds to weak coupling, see Eq. (7) of Ref.~\cite{Sch88}. 
The solution shows that in 2D or 3D 
a single orbiton in the undoped AF system described by the $t$--$J$ model 
cannot fractionalize due the magnetic string effect. Still, the orbiton
dressed with spin fluctuations is mobile on a renormalized scale~\cite{Sch88}.
For example the 2D case directly corresponds to the result 
in Ref.~\cite{Mar91} which gives $\omega \approx E_z -x_1 + x_2 (\cos k_x + \cos k_y)^2
+x_3 (\cos 2k_x + \cos 2 k_y)$ for the orbiton dispersion ($x_i$ are positive parameters 
$\propto t^2/J$ when $J > t$, cf. Ref. \cite{Mar91} for exact values).
We thus expect to observe, e.g. in high resolution RIXS experiment on 
2D cuprates, a spectrum similar to the one-particle Greens function of the higher-dimensional $t$--$J$ model, 
with a low-energy quasi-particle orbiton peak $\propto \omega$ and additional incoherent 
part~\cite{Mar91}. Crucially, while the parameters $x_i$ depend on $J/t$ and would vary in 
more realistic spin-orbital models \cite{note}, the general shape of the orbiton 
dispersion $\omega$ -- the 
minimum at $(\pi /2, \pi/2)$, saddle point at $(\pi, 0)$, maxima at 
$(0,0)$ and $(\pi,\pi)$ -- is robust reflecting the fact that 
the coherent motion of orbiton is possible only within a given spin-sublattice. 

{\it No analogue in spin sector.---} 
Let us now revisit the analogue where the spin is polarized (FM) and the
orbital sector shows AO order
due to, e.g., a large Jahn-Teller effect. 
An example would be the FM and AO planes in LaMnO$_3$ or 
KCuF$_3$~\cite{Mou96}.
Many studies have shown that the orbital degrees 
of freedom merely renormalize the spin excitation in this case and do not 
change the periodicity of magnons (see, e.g., Ref.~\cite{Khaliullin:2003p2580}).
This qualitative difference is caused by the fact that in realistic cases the
Jahn-Teller stabilized AO order is much more classical and robust, thus
suppressing the creation of orbital excitations by the excited spin and 
giving larger weight to the pure spin excitation. In other words, spin waves 
are typically below the orbital gap and well protected by the underlying 
$SU(2)$ symmetry and Goldstone theorem. 

In conclusion, we have shown that orbitons in realistic spin-orbital models 
with AF-spin and ferroorbital ground state are so strongly coupled to the 
spin excitations that the usual mean-field decoupling of two sectors breaks down. In fact, 
we have presented an exact mapping of the problem onto an effective $t$--$J$ 
model and have shown that the orbiton in such models behaves like a single 
hole in undoped AF Mott insulator. However, since the typical superexchange 
parameters for KK spin-orbital models lead to $J>t$, the study of orbiton 
problem provides access to a regime of the $t$--$J$ model which has been 
thought to be `unphysical' in terms of single-band Hubbard models. 
Finally, in 1D signatures of spin-orbital separation are expected again
in the peculiar regime where spinons are faster than orbitons. 


We thank G. Jackeli, M.W. Haverkort and A.M. Ole\'s for discussions. Support 
from the Alexander von Humboldt Foundation (K.W.) and the DFG Emmy Noether 
Program (M.D.) is acknowledged.


\section{Supplementary Material}
In what follows we show that the $t$--$J$ model description of the orbiton
problem [Eqs. (6-7) in the main text] is not only valid in the 1D case
but also in higher dimensions. To be explicit we concentrate now on
the 2D case (from which the 3D case follows in a straightforward way)
and introduce the Jordan-Wigner fermions $\alpha$ and $\beta$ for pseudospins and spins:
\begin{align}
S^+_j  &= \beta_j  \e^{  i \pi Q_j }, \\
S^-_j  &=  \e^{ - i \pi Q_j } \beta^\dag_j, \\
S^z_j &= \frac12 - n_{j \beta}, 
\end{align}
where $Q_j= \sum_{l=1}^{j-1} n_{l \beta}$ and $\beta^\dag_j$ create spinons while 
\begin{align}
T^+_j &= \e^{ - i \pi \bar{Q}_j } \alpha^\dag_j, \\
T^-_j  &= \alpha_j \e^{  i \pi  \bar{Q}_j}, \\
T^z_j  &= n_{j \alpha} - \frac12,
\end{align}
where $\bar{Q}_j = \sum_{l=1}^{j-1} n_{l \alpha}$. Note that this is the same
transformation as in the main text but, to keep track of the phase factors, 
we explicitly wrote the site index $j$ of the phase factors $Q_j$ and $\bar{Q}_j$. 

Next, similarly as in 1D, the spin of the propagating electron in the upper 
orbital 2 is conserved, and one may calculate the spectral  
function $O(k, \omega)$ for, e.g., spin-up in the upper orbital. 
We are then allowed to replace 
$T^{\dag}_k \rightarrow \sum_{j} \e^{ikj} T^+_j (\frac12 + S^z_j) = 
\e^{ - i \pi \bar{Q}_j }\alpha^\dag_j (1- n_{j \beta})$ in Eq.~(2) in the main text
and thus terms in the Hamiltonian that create or annihilate both a spinon and a
pseudospinon at the same site also lead to a vanishing contribution to the
spectral function. 

In 2D one has to take care of the phase factors. However, the crucial observation
is that for the case of the {\it single} orbiton the phase factors $\bar{Q}_j$ associated
with a pseudospinon either do not contribute at all [Eq. (\ref{eq:spectralpl}) below] or 
cancel for the nearest neighbor bonds [Eq. (\ref{eq:eff}) below] -- similarly to 1D. Thus, these are only the spin phase
factors $Q_j$ which are in the end present in the spectral function and
Hamiltonian written in terms of the spinons and pseudospinons:
\begin{align} \label{eq:spectralpl}
O(k, \omega)=\frac{1}{\pi} \lim_{\eta \rightarrow 0} 
\Im \langle \bar{\psi} | \alpha_k \frac{1}{\omega \! +\! E_{\bar{\psi}}  \!-\!
\mathcal{\bar{H}}\! -\! E_z \! -\! i \eta } \alpha^+_k | \bar{\psi} \rangle, 
\end{align}
with the effective fermionic Hamiltonian 
\begin{align} \label{eq:eff}
\mathcal{{\bar{H}}}\! =\! -\frac12  J\! \sum_{\langle i, j \rangle}&
\big(\e^{ - i \pi  {Q}_i} \beta^\dag_i\alpha_i\alpha^\dag_{j} \beta_{j}\e^{  i \pi  {Q}_j} \! +\! \alpha_i
\alpha^\dag_{j} \!+\! h.c.\big)\! 
\nonumber \\ 
\!+\! J\! \sum_{\langle i, j \rangle} &\Big[ \frac12 (\e^{ - i \pi  {Q}_i}\beta^\dag_i \beta_{j} \e^{  i \pi  {Q}_j}\!+\! h.c.)  \nonumber \\
&\!-\! \frac12 (n_{i \beta}\! +\! n_{j \beta} )  \! +\! n_{i \beta} n_{j \beta} \Big].
\end{align}
As stated above (and mentioned for the 1D case in the main text), here we have 
an implicit constraint $\forall_j \  \beta^\dag_j \beta_j + \alpha^\dag_j
\alpha_j \leq 1$ while $|\bar{\psi}\rangle$ is a tensor product of the 
magnetic ground state $|\psi_S\rangle=|\textrm{AF}\rangle$ expressed in 
terms of spinons and a vacuum state for pseudospinons.

At this point, we observe that the resulting effective Hamiltonian is in fact
a Hamiltonian for the $t$--$J$ model written in terms of the Jordan-Wigner
fermions with the above constraint. By introducing
the electron operators 
${p}_{j \uparrow} = \alpha^\dag_j$, $\ {p}_{j \downarrow} = \alpha^\dag_j
\beta_j \e^ { i \pi Q_j }$ 
acting in the restricted Hilbert space without double occupancies we obtain 
\begin{align} \label{eq:spectraltJ}
O(k, \omega)\!=\!\frac{1}{\pi} \lim_{\eta \rightarrow 0} 
\Im \langle {\tilde{\psi}}  |{p}^{\dag}_{k \uparrow}   \frac{1}{\omega \!+\!
E_{{\tilde{\psi}} }  \!-\! \mathcal{{\tilde{H}}} \!-\! E_z \!-\! i \eta } {p}_{k
\uparrow}  |{\tilde{\psi}} \rangle, 
\end{align}
with the $t$--$J$ Hamiltonian 
\begin{align} \label{eq:tJ}
{\mathcal {\tilde{H}}}\! = &\! -\! t \!\sum_{\langle i, j \rangle, \sigma} ({p}^\dag_{i \sigma}
{p}_{j \sigma}\! +\! h.c.) 
\! +\! J \sum_{\langle i, j \rangle } ({\bf S}_{i } \cdot {\bf S}_{j} \! +\! \frac14 {n}_{i}
{n}_{j}), 
\end{align}
where ${n}_{j } =\sum_{\sigma} n_{jp\sigma}$ and the hopping parameter $t$ is 
defined as $t=J/2$. The ground state $|\tilde{\psi}\rangle$ is now
the tensor product of a vacuum state for holes and the $|\psi_S\rangle$ state. 
We have thus mapped the problem of a {\it single orbital excitation} in the 2D FO and AF ground state, 
with dynamics governed by Hamiltonian Eq. (1) in the main text, onto a {\it single hole} doped 
into the undoped 2D AF ground state with its dynamics governed by Hamiltonian 
(\ref{eq:tJ}).


\begin{thebibliography}{10}
\bibitem{note2} It turns out that the results below are valid 
also when the symmetry of the orbital superexchange 
term is further broken (as it may be the case in some compounds).
\bibitem{note}
Note that inequivalent hoppings of orbitals [not included in model 
Eq.~(\ref{eq:h})] would lead to $t \equiv 2t_1t_2/U$, $J=4t^2_1/U$ and 
$t \neq J/2$, where $t_1$ ($t_2$) is the hopping of electrons in the
lower (upper) orbital. Still $J>t$ holds since $t_1 \geq t_2$.
\end{thebibliography}
\end{document}